\documentclass[aps,twocolumn,superscriptaddress,altaffilletter,lengthcheck, tightenlines,showpacs,showkeys]{revtex4}
\usepackage{graphicx}
\usepackage{bm}

\newcommand{\be}{\begin{equation}}
\newcommand{\ee}{\end{equation}}
\newcommand{\ben}{\begin{eqnarray}}
\newcommand{\een}{\end{eqnarray}}
\newcommand{\n}{\label}
\newcommand{\no}{\noindent}

\newcommand{\om}{\Omega}
\newcommand{\bi}[1]{\mbox{\boldmath$#1$}}

\date{\today}

\begin{document}
\author{L.P. Chimento}

\author{M\'onica Forte}

\affiliation{Dpto. de F\'\i sica, Facultad
de Ciencias Exactas y Naturales, Universidad de Buenos Aires,
Ciudad Universitaria, Pabell\'on I, 1428 Buenos Aires,Argentina}

\title{Unified model of Baryonic matter and dark components}

\begin{abstract} 
	We investigate an interacting two-fluid cosmological model and
 introduce a scalar field representation by means of a linear combination of
 the individual energy densities. Applying  the integrability condition to
 the scalar field equation we show that this "exotic quintessence" is driven
 by an exponential potential and the two-fluid mixture can be considered as
 a model of three components. These components are associated with baryonic
 matter, dark matter and dark energy respectively. We use the $Simon,
 Verde \ \& \ Jimenez (2005)$ determination of the redshift dependence of
 the Hubble parameter to constrain the current density parameters of this
 model. With the best fit density parameters we obtain the transition
 redshift between non accelerated and accelerated regimes $z_{acc}=0.66$ and
 the time elapsed since the initial singularity $t_0= 19.8 Gyr$. We study
 the perturbation evolution of this model and find that the energy density
 perturbation decreases with the cosmological time.
	
\end{abstract}


\maketitle

\section{Introduction}

	Astrophysical data suggests that the Universe is accelerating \cite{Perlmutter:1998np,Spergel:2003cb}.  This acceleration may be explained by very different models, among them, the simplest one is the $\Lambda$CDM \cite{Padmanabhan:2002ji,Sahni:1999gb}. It assumes a cosmological constant arising from the energy density of the zero point fluctuations of the quantum vacuum and cold dark matter in 
form of pressureless dust. While it fits rather well all the observational constraints, the small positive value of the energy density of the vacuum remains as an explanatory challenge for  physics today. See also Refs. \cite{Triay:2005jr,Mersini-Houghton:2006ue} for objections to this interpretation. The next step is to propose a dark energy component that may vary with time and that is generally modeled by a scalar field. 
Most of these models assume that  dark matter and  scalar
field components evolve independently.
Again, this is not the solution because in the analysis of these models through SNIa or WMAP data, best fit models for one set data alone is usually ruled out by the other set at a large confidence limit \cite{Jassal:2005qc}.  As these conclusions are valid only for standard models, where dark energy and dark matter are decoupled, many papers have been devoted to interacting models \cite{Amendola:2006ku}. Certain models conceive the interaction as a time-variable dark mass, evolving with an inverse power law potential or an exponential potential, \cite{Farrar:2003uw}, \cite{Amendola:2002kd}, \cite{Neupane:2007mu}, 
\cite{Axenides:2004kb}. 

In this paper we clarify this point and establish an exactly
solvable model with a smooth transition from a matter dominated phase
to a period of accelerated expansion. We introduce an interacting two-fluid cosmological model and investigate the effects of imposing the integrability condition of the whole equation of conservation. It forces the dynamic of the model to be  governed by a modified Friedmann equation with three components. One of them, associated with an exponential potential, drives an exotic scalar field, (exotic quintessence). Here, we show that the problem of an accelerating universe can be realized in a comparatively simple manner within the framework of general relativity. Finally, the perturbation evolution of the model is investigated.

Our paper is organized as follows. In section II we consider the general interacting two-fluid cosmological model and introduce the exotic quintessence. There, we obtain the evolution equation of the exotic field, find their implicit solutions, build the modified Friedmann equation and show the asymptotic behavior of the scale factor by using stability analysis. In section III we introduce baryonic matter, dark matter and dark energy components and show that dark components satisfy separately an effective equation of conservation with variable equation of state. In section IV we find confidence regions for the parameters of the model by using the Hubble function H(z) data, the age of universe and the redshift of the transition from non-accelerated to accelerated regime for the best fit model. In section V we present the equations governing the perturbations of the model. In section VI we express the conclusions.

\section{Exotic quintessence}

Cosmological models are based in the Einstein equations of the gravitational field where the source includes different kinds of matter known, for instance protons, neutrons, photons, neutrinos, etc., as well as non-relativistic non-baryonic cold dark matter and dark energy. 
We embark in a less ambitious project by considering a model consisting of two perfect fluids with an energy momentum-tensor $T_{ik} =
T_{ik}^{(1)}+T_{ik}^{(2)}$. Here $T_{ik}^{(n)} = (\rho_{n}+p_{n})
u_{i}u_{k}+p_{n}g_{ik}$, where $\rho_{n}$ and $p_{n}$ are the energy
density and the equilibrium pressure of fluid $n$ and 
$u^{i}$ is the four-velocity. Assuming that the two fluids interact between them in a spatially flat homogeneous and isotropic Friedmann-Robertson-Walker
(FRW) cosmological model, the Einstein's equations reduce to two algebraically independent equations:
\ben
\n{00}
3 H^2 =\rho_1 +\rho_2, \\
\n{c}
\dot\rho_1+\dot \rho_2 + 3H((1+w_1)\rho_1 + (1+w_2)\rho_2)= 0,
\een

\no where a(t) is the FRW scale factor and $H(t)=\dot a/a$ is the Hubble expansion rate. We introduce an equation of state for each fluid component $w_n=p_{n}/\rho_{n} \ \ n = 1,2$ and for simplicity we assume that $w_1,w_2$ are constants and $\rho_1,\rho_2>0$. This simplified model leads to a reduction of the number of fundamental parameters required to describe observations. It can be considered as an advantage from the computational point of view. We choose units such as the gravitational constant is set to $8\pi G=1$ and $c=1$.

The whole equation of conservation (\ref{c}) shows the interaction between both fluid components allowing the mutual exchange of energy and momentum, meaning that, there will be no local energy-momentum conservation for these fluids separately. Then, we assume an overall perfect fluid description with an effective equation of state, $w=p/\rho=-2\dot H/3H^2 -1$, where $p=p_1+p_2$ and $\rho=\rho_1+\rho_2$. So that, from Eqs. (\ref{00})-(\ref{c}) we get
\be
\n{h.}
-2\dot H=(1+w_1)\rho_1+(1+w_2)\rho_2=(1+w)\rho.
\ee

\no To avoid an eventual super acceleration of the universe, which could lead to a ``big rip" singularity, we choose $w_1>-1$ and $w_2>-1$ \cite{crossing}. Hence,  $(1+w)\rho=-2\dot H>0$ and there is no big rip singularity. These models can be investigated by introducing a scalar field $\phi$ representation of the interacting two-fluid mixture 
\be
\n{f2}
\dot\phi^2=(1+w_1)\rho_1+(1+w_2)\rho_2,
\ee
with $\dot\phi^2=-2\dot H$. The dynamic equation for the scalar field is obtained from the equation of conservation (\ref{c})
\be
\n{kg}
\ddot\phi + \frac{3}{2}(1+w_1) H\dot\phi+ \frac{w_1 - w_2}{2}\,\frac{\dot\rho_2}{\dot\phi} = 0.
\ee

\no It can be integrated by setting the interaction between the two fluids by 
\be
\n{c2}
\dot\rho_2 + A\dot\phi \rho_2 = 0,
\ee
where $A$ is a new constant parameter of the model. Integrating Eq. (\ref{c2}), we find that the energy density of the second fluid can be associated with an exponential potential
\be
\n{v}
\rho_2=\rho_{20}H_0^2e^{-A(\phi-\phi_0)}=V(\phi),
\ee
where $\rho_{20}$ is a positive integration constant and $H_0$, $\phi_0$ are the present values of Hubble constant and scalar field. 

	From Eqs. (\ref{f2}), (\ref{kg}) and (\ref{v}) we obtain the total energy density and pressure of the fluid mixture and the dynamical equation for the scalar field
\ben
\n{r}
\rho=\frac{\dot\phi^2}{1+w_1}+\frac{w_1-w_2}{1+w_1}\,\,V,\\
p=w_1\frac{\dot\phi^2}{1+w_1}-\frac{w_1-w_2}{1+w_1}\,\,V,\\
\ddot\phi + \frac{3}{2}(1+w_1) H\dot\phi+ \frac{w_1 - w_2}{2}\,\frac{dV}{d\phi} = 0.
\een

\no with $\rho+p=\dot\phi^2$. These equations are different than the conventional ones describing quintessence, in contrast, define an exotic scalar field. When the interacting two-fluids system is related 
to the scalar field in the form $\rho_{1}=\dot\phi^2/2$ and $\rho_{2}=V(\phi)$, with equations of state $p_{1}=\rho_{1}$ and $p_{2}=-\rho_{2}$, meaning that $w_1=1$ {\it (stiff matter)} and $w_2=-1$ {\it (vacuum energy)}, the exotic scalar field reduces to quintessence. Then, due to the interactions between the two-fluid components the energy-momentum tensor conservation of the system, as a whole, is equivalent to the Klein-Gordon equation. For any other interacting two-fluid mixture, the cosmological model contains an exotic quintessence field $\phi$ driven by an exponential potential. 

Using the integrability condition (\ref{c2}) in the field equation (\ref{kg}), its first integral is given by  
\be
\n{f}
\dot\phi=AH+cH_0(1+z)^{3(1+w_1)/2},
\ee 
and
\be
\n{fi}
\phi=\phi_0-A\ln{(1+z)}-cH_0\int^z_{0}\frac{(1+z)^{(1+3w_1)/2}}{H}\,{dz},
\ee

\no where $c$ is an arbitrary integration constant, $z=-1+a_0/a$ is the redshift
 parameter and $a_0$ is the present scale factor.
 This model is finally closed when the Eq. (\ref{f}) is inserted into the
 energy density (\ref{r}) \cite{2c} .
 Hence, the Friedmann equation (\ref{00}) reads

\ben\nonumber
3H^2 = \frac{1}{3(1+w_1)-A^2} \left[6cAH_0H(1+z)^{3(1+w_1)/2}+ \right.
\\
\left.
\n{F}
3(w_1-w_2)\rho_2+3c^2H_0^2(1+z)^{3(1+w_1)} \right]
\een

\no As a consequence of the linear term in the expansion rate $H$, this equation can be seen as amodified Friedmann equation. Its solution gives the scale factor and the model we propose, containing exotic quintessence, could be formally solved.

In order to obtain the asymptotic behavior of the scale factor it will be
 useful to find the constant solutions of the dynamical equation for the
overall equation of state 
\be
\n{.gt}
\dot w=-3H(w-w_1)\left(\sqrt{\frac{A^2}{3}\ (1+w)}-(1+w)\right),
\ee

\no and investigate their asymptotic stability. This equation has two stationary solutions $w_1 $ and $A^2/3 - 1$ (Eq. (\ref{h.}) excludes the solution $w=-1$).
Assuming that $A^2<3(1+w_1)$, we find that $w_1$
is an unstable solution while $A^2/3 -1$
becomes asymptotically stable.
Essentially, the evolution of the geometry is dictated by
$w=-2\dot H/3H^2 - 1$, meaning that the universe begins
to evolve from an unstable phase as it were dominated by the first
fluid $w_1$ at early times, $a\approx t^{2/3(1+w_1)}$, and
ends in a stable expanding phase dominated by the exponential potential,
$a\approx t^{2/A^2}$. The latter becomes an expanding accelerated phase
when the slope of the potential satisfy the inequality $A^2<2$.

\section{Dynamic of baryonic and dark components}

The integrability condition (\ref{c2}) can be considered as an effective equation of conservation for the second fluid. This allows us to identify $\rho_2$ with the energy density of the dark energy component and $w_{de}= -1 + A\dot\phi/3H$ with its effective equation of state. 
Expressing the latter in term of the exotic field $\phi$ we get
\be
\n{gdef}
w_{de}\equiv -1 + \frac{A\dot\phi}{3H}= -1 + \frac{A\dot\phi}{\sqrt{3[\dot\phi^2-w_2V]}},
\ee
and the relation
\be
\n{gt}
w= -1 + \frac{3}{A^2}(1+w_{de})^2,
\ee

\no linking the overall and dark energy equations of state. Also, for convenience we  write $\rho_2=\rho_{20}H_0^2(1+z)^{3\lambda}$ with 
\be
\n{gaf}
3\lambda\ln{(1+z)}=-A(\phi-\phi_0).
\ee

\no The $H$ linear term in the Eq. (\ref{F}) is adequate to describe non-relativistic non-baryonic cold dark matter components 
whose energy-momentum tensor is approximately dust-like. Finally, the baryonic matter is introduced by setting $w_1=0$ in the third term of the Eq. (\ref{F}). Making these identifications the modified Friedmann equation (\ref{F}) becomes
\be
\n{00m}
\frac{H^2}{H_0^2}=\om_{dm}(1+z)^{3/2}\frac{H}{H_0}+\om_{de}(1+z)^{3\lambda}+\om_b(1+z)^{3},
\ee
where
\ben
\n{odm}
\om_{dm}=\frac{2cA}{3-A^2},\\
\om_{de}=\frac{-w_2\rho_{20}}{3-A^2},\\
\n{ob}
\om_b=\frac{c^2}{3-A^2},
\een

\no are the present dark matter, dark energy and baryonic matter density parameters respectively. As these are constrained according to 
$\om_{dm}+\om_{dm}+\om_b=1$, we conclude that
\be
\n{rel}
(c+A)^2-w_2\rho_{20}=3.
\ee

\no During the accelerated epoch of the universe $\ddot a>0$, the SEC is
violated and $\rho+3p=\rho_1+(1+3w_2)\rho_2<0$ leads to
$-1<w_2<-1/3$. Hence, one finds that the conditions $A^2<3$ and $cA>0$
are consistent with the requirement of having positive density parameters.

The exotic quintessence is essentially based on the integrability condition
(\ref{c2}) under which the conservation equation (\ref{c}) and the exotic
field equation (\ref{kg}) can be integrated. Actually, from the above
condition $cA>0$ along with Eq. (\ref{f}), we get $A\dot\phi>0$ for an
expanding universe. Then, $\rho_2$ is a Liapunov function and the solution
$\rho_2=0$, of Eq. (\ref{c2}), is asymptotically stable.  Also, from Eqs.
(\ref{f2}), (\ref{v}) and (\ref{f}) the energy density of the first fluid
has a vanishing limit in the remote future. Hence, this general model is
viable and it does not contradict basic cosmological conjectures.
 
  From Eqs. (\ref{odm}) and (\ref{ob}) we can express the parameter $A$ and the integration constant $c$ in terms of the present density parameters $\om_{dm}$,
\ben
\n{Ac}
A = \frac{\pm\ \sqrt{3}\ \om_{dm}}{\sqrt{\om_{dm}^2+4\om_b}},\\
c=\frac{\pm\  2\sqrt{3}\ \om_{b}}{\sqrt{\om_{dm}^2+4\om_b}}
\een
besides,
\be
\n{r02}
-w_2\rho_{20}=\frac{12\om_{de}\om_b }{\om_{dm}^2+4\om_b}.
\ee

Finally the original problem of the interacting two-fluid mixture governed by system equations (\ref{00})-(\ref{c}) is equivalent to an effective model with a ``three-fluid" mixture. So that, the effective dynamical equations of our model read
\ben
\n{00'}
3H^2=\rho_{dm}+\rho_{de}+\rho_b,\\
\dot\rho_{dm}+3H(1+w_{dm})\rho_{dm}=0,\\
\n{cde}
\dot\rho_{de}+3H(1+w_{de})\rho_{de}=0,\\
\dot\rho_{b}+3H\rho_{b}=0,
\een
where
\ben
\rho_{dm}=3H_0\om_{dm}(1+z)^{3/2}H,\\
\rho_{de}=3H_0^2\om_{de}(1+z)^{3(1+\lambda)},\\
\rho_{b}=3H_0^2\om_{b}(1+z)^{3},
\een
are the effective energy densities of dark and baryonic components and 
\be
\n{gde}
w_{de}=\frac{2\om_{b}}{\om_{dm}^2+4\om_b}\left[-2+
\frac{\om_{dm}H_0}{H}(1+z)^{3/2}\right],
\ee
is the effective equation of state of the dark energy. Also, we find the following relation
\be
\n{gdm}
w_{dm}=\frac{1}{2}\left[-1+\frac{3}{A^2}(1+w_{de})^2\right],\\
\ee

\no between the effective equations of state of dark components. So that, the
knowledge of them determines the remaining ones including $w$.
To obtain the above effective dynamical equations of the model we have taken
into account that $\rho_{de}\propto\rho_2$. Hence, we have identified the
equation of conservation (\ref{c2}) with (\ref{cde}) to express, after using
(\ref{r}), the effective $w_{de}$ for dark energy in terms of the
observed density parameters, the present Hubble expansion rate and the
redshift parameter.

In our model the stationary solutions $w^e=0$ at early times and $w^l=w_{de}^l= - 1 + A^2/3$ at late times along with Eqs. (\ref{gt}) and (\ref{gdm}) lead to the stationary solutions $w_{de}^e= - 1 + A/\sqrt{3}$, $w_{de}^l= - 1 + A^2/3$ and $w_{dm}^e=0$, $w_{dm}^l=(-1+A^2/3)/2$. As the observed density parameter satisfies the condition $\om_{dm}^2<8\om_b$, then $A^2<2$ by using Eq. (\ref{Ac}). In this case, we find that $w_{de}^e$ is an unstable solution at early times and $w_{de}^l$ becomes asymptotically stable at late times. Here, the evolution of the geometry represents a universe that begins to evolve as it were matter dominated at early times and ends in an accelerated phase dominated by the dark energy component. As the accelerated epoch begins at $\ddot a=0$ or $w=-1/3$, the corresponding redshift is given by the expression
\be
\n{za}
z_{acc}=-1+\left[\frac{\om_b}{\om_{de}(\sqrt{2}-A)^2}\right]^{w(z_{acc})/3},
\ee

\no On the other hand, the Eq. (\ref{gde}) allow us to find the elapsed time $t_0$ from the creation 
\be
\n{t0}
t_0=\frac{\om_{dm}}{3\om_{b}H_0}\left[\frac{\sqrt{3}}{A}-1\right],
\ee
where we have used that the scale factor behaves as $a\approx a_0(t/t_0)^{2/3}$, at early times.
 
\section{Observational Constraints}

	In Ref.  \cite{Lazkoz:2007zk} it was used the recently published Hubble function $H(z)$ data [SV\&J(2005)]\cite{Simon:2004tf}, extracted from differential ages of passively evolving galaxies. This is interesting for, among other reasons, the function is not integrated over, in contrast to standard candle luminosity distances or standard ruler angular diameter distances 
	Since the Hubble parameter depends on the differential age of the Universe as a function of z in the form $H(z)= - (1+z)^{-1}dz/dt$, it can be directly measured through a determination of dz/dt. In the procedure of calculating the differential ages, Simon et al. have employed the new released Gemini Deep Deep Survey \cite{Abraham:2004ra} and archival data \cite{Treu:2001hq}, \cite{Nolan:2003bt}
to determine the  9 numerical values of $H(z)$ in the range $0<z<1.8$,  and their errors [see Table 1]. These data will be inserted in our Eqs. (\ref{gaf}) and (\ref{00m}) to derive restrictions on the range of possible values for the density parameters (\ref{odm})-(\ref{ob}).	
\begin{table}[htbp]
   \begin{center}
\begin{tabular}{r|c|c}
\hline
$z$&$H(z)$&$1\sigma$\\
$ $&$(km s^{-1}Mpc^{-1})$&$uncertainty$\\
\hline
0.09&69&$\pm12$\\
0.17&83&$\pm8.3$\\
0.27&70&$\pm14$\\
0.40&87&$\pm17.4$\\
0.88&117&$\pm23.4$\\
1.30&168&$\pm13.4$\\
1.43&177&$\pm14.2$\\
1.53&140&$\pm14$\\
1.75&202&$\pm40.4$\\
\hline
\end{tabular}
 \end{center}
    \caption{\,SV\&J Hubble Parameter vs. Redshift Data.}
    \label{tab:Example}
\end{table}

We adopt  the prior $H_0 = 72 km s^{-1}Mpc^{-1}$ for $H(z=0)$. It is exactly the mean value of the results from the Hubble Space Telescope key project \cite{Freedman:2000cf} and consistent with the one from WMAP 3-year result \cite{Spergel:2006hy}. 

The parameters of the model can be determined by minimizing the function
\be
\n{ch1}
\chi^2(\Omega_{dm},\Omega_{de}) = \sum_{i=1}^9 \frac{[H_{th}(\Omega_{dm},\Omega_{de}; z_i) - H_{ob}(z_i)]^2}{\sigma^2(z_i)}
\ee
where
$$
H_{th}(\Omega_{dm},\Omega_{de}; z_i)= H_0 (1+z_i)^{3/2}\ \  \times$$
\be
\n{ch2}
\left[\frac{\Omega_{dm}}{2} +\sqrt{\left(1-\frac{\om_{dm}}{2}\right)^2 +
\om_{de}\left[ (1+z_i)^{3(\lambda-1)}-1\right]}\right]
\ee

\no is the predicted value for the Hubble parameter, obtained from Eq. (\ref{00m}), and $\lambda$ is calculated from Eqs. (\ref{fi}) and (\ref{gaf}).

\no $H_{ob}$ is the observed value of $H$ at the redshift $z_i$, $\sigma^2$ is the corresponding $1\sigma$ uncertainty, and the summation is over the 9 observational $H(z_i)$ data points at redshift $z_i$ \cite{Wei:2006ut}. Also, we adopt the prior $\Omega_b=0.05$ \cite{Astier:2005qq}.

\begin{figure}
\centering 
\begin{minipage}{0.89\linewidth}
\includegraphics[width=8.0cm]{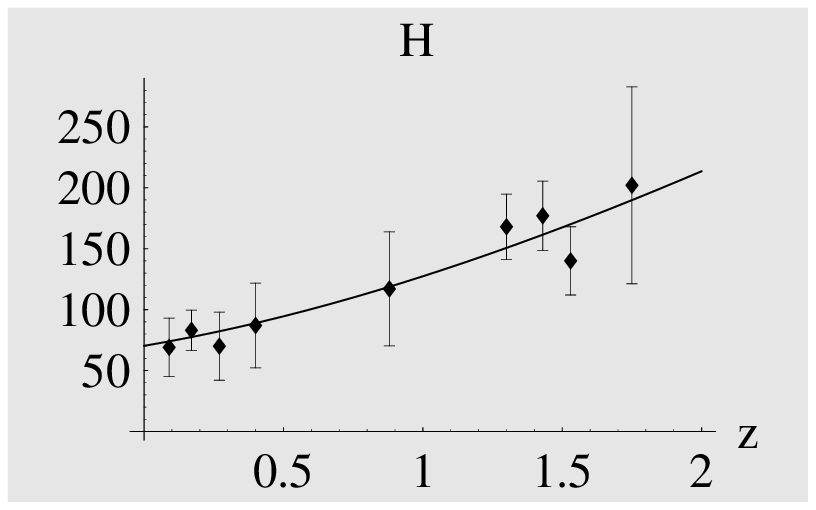}
\center{Fig. 1: Observational $H(z)$ with $1\sigma$ uncertainties from $SV\&J(2005)$ and the best fit theoretical $H(z)$.}
\end{minipage}
\end{figure}

\begin{figure}
\centering 
\begin{minipage}{0.79\linewidth}
\includegraphics[width=7.5cm]{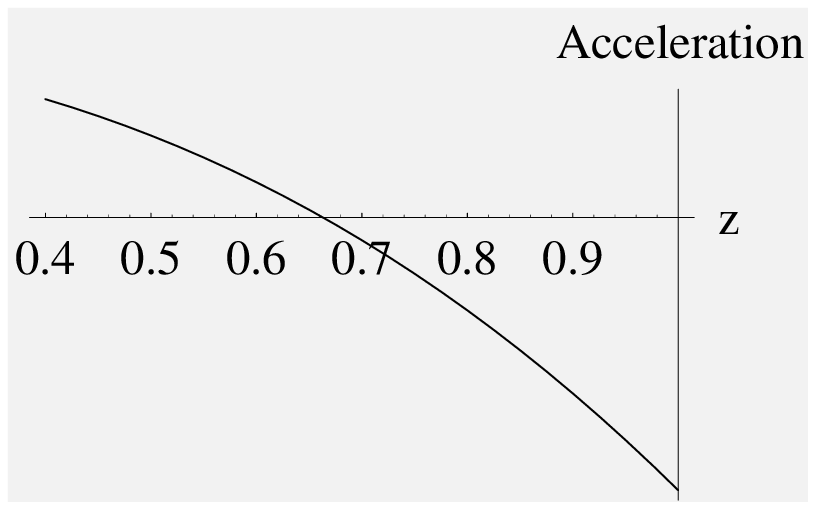}
\center{Fig. 2: Acceleration vs. redshift for the best fit model.}
\end{minipage}
\end{figure}

\begin{figure}
\centering 
\begin{minipage}{0.89\linewidth}
\includegraphics[width=8.0cm]{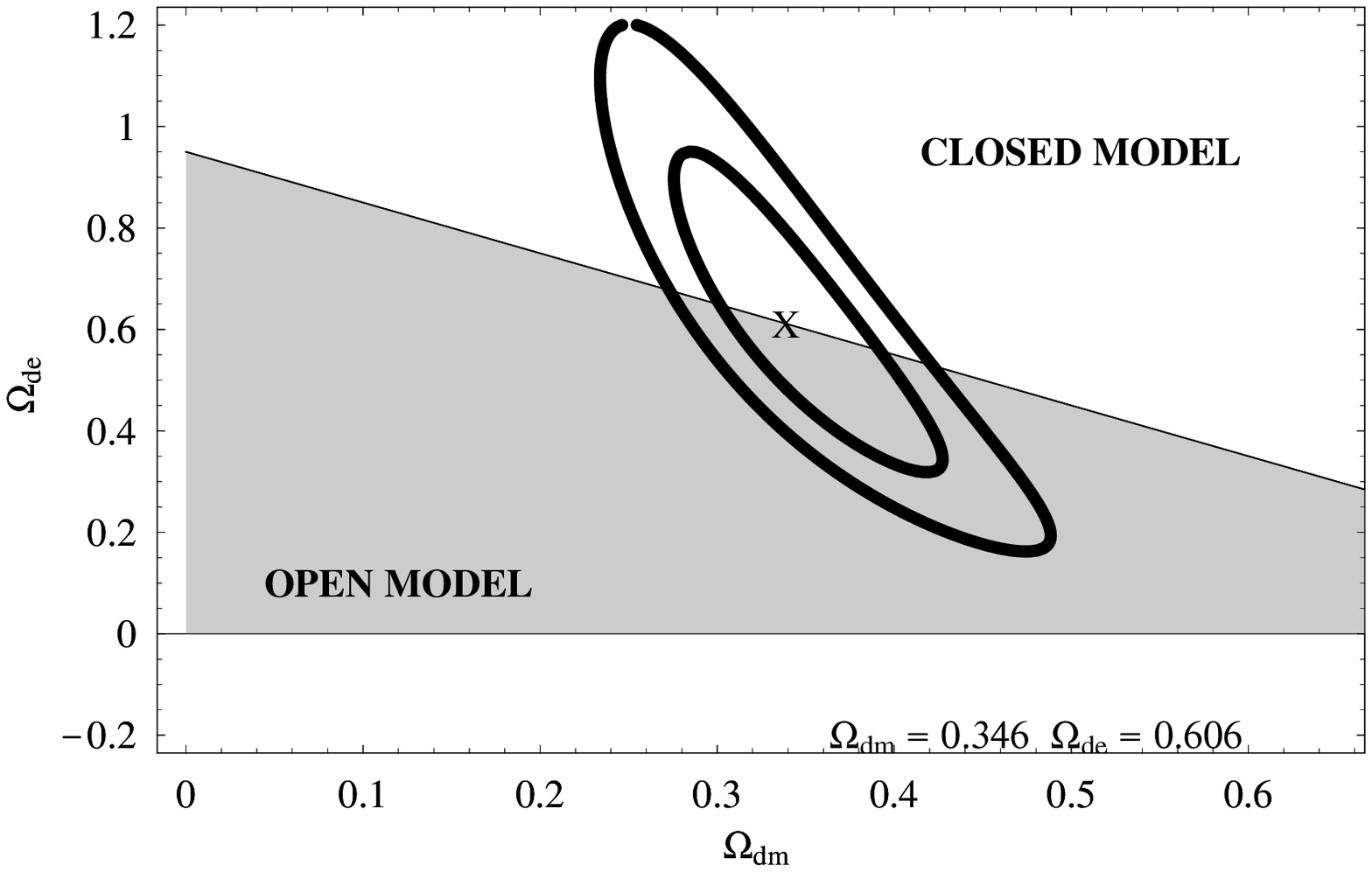}
\center{Fig. 3: The $1\sigma$ and $2\sigma$ confidence regions, (inside and between the elliptic contours) for $\Omega_{dm}$ and $\Omega_{de}$ from $SV\&J(2005)$.  The cross is the best fit model. The  straight line corresponds to the flat cosmology making the separation between  open and closed universes.}
\end{minipage}
\end{figure}

\begin{figure}
\centering 
\begin{minipage}{0.89\linewidth}
\includegraphics[width=7.5cm]{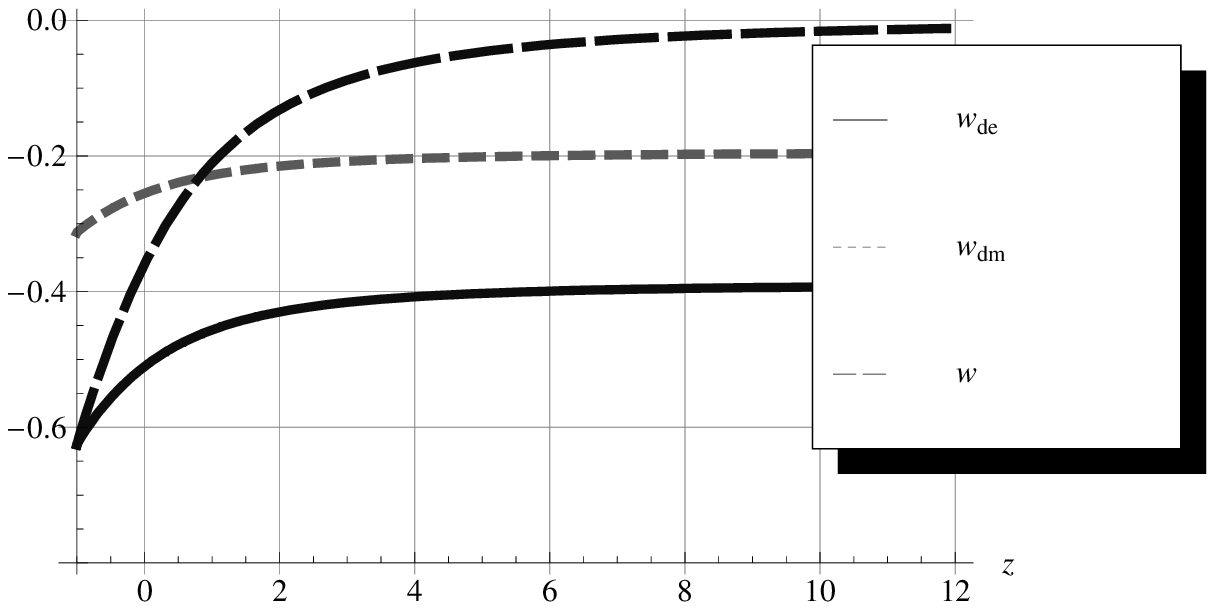}
\center{Fig. 4: State parameters for dark matter, dark energy and overall fluid.}
\end{minipage}
\end{figure}

\begin{figure}
\centering
\begin{minipage}{0.89\linewidth}
\includegraphics[width=8.0cm]{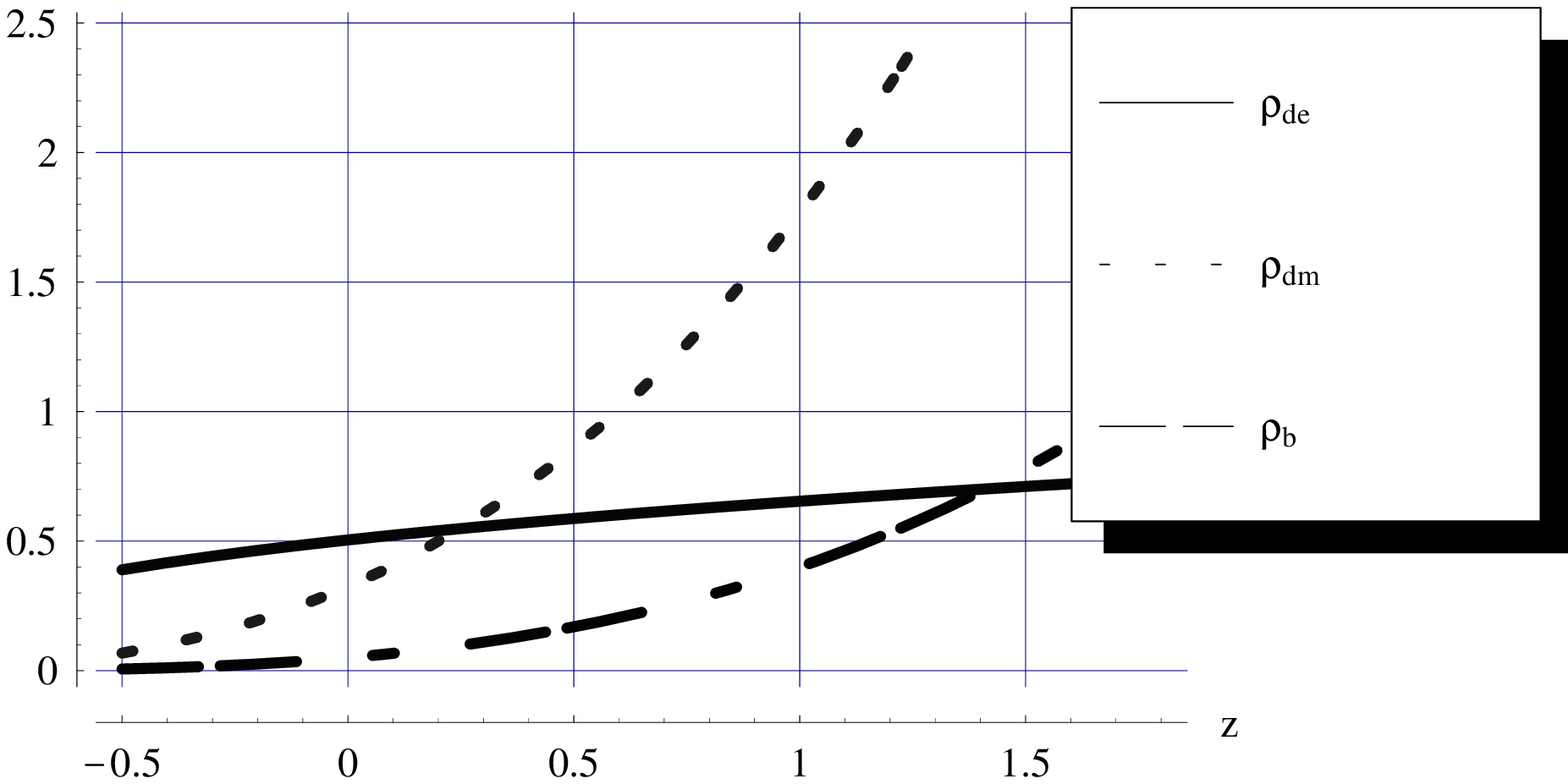}
\center{Fig. 5: Energy densities (in units of $3H_0^2$) vs. redshift z}
\end{minipage}
\end{figure}

\begin{figure}   
\centering 
\begin{minipage}{0.89\linewidth}
\includegraphics[width=8.0cm]{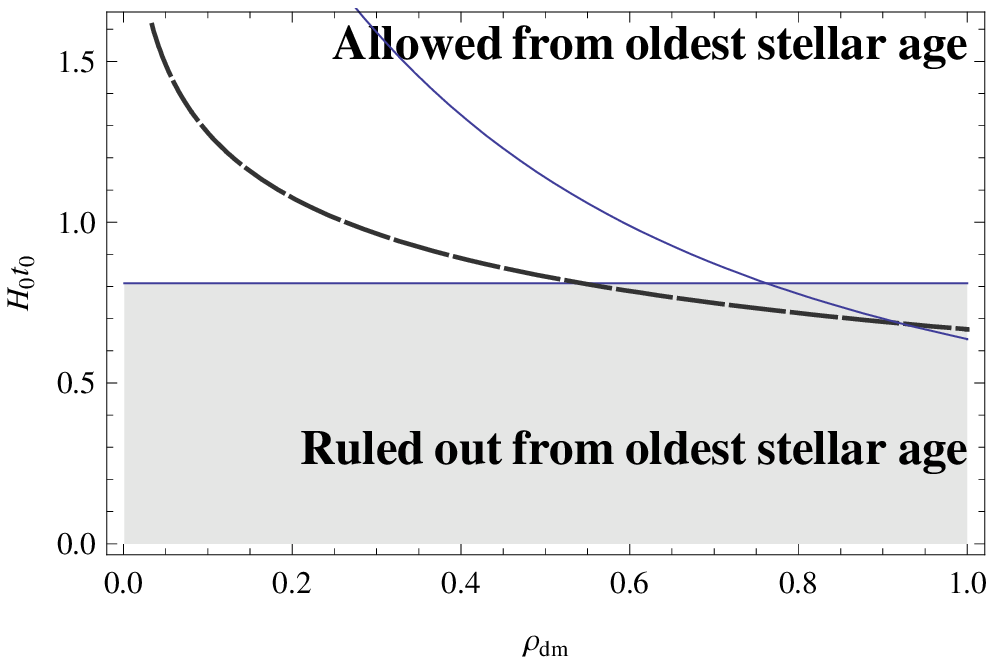}
\center{Fig. 6: The age of the universe (in units of $H_0^{-1}$). The solid line corresponds to our model and the dashed line represents the $\Lambda$CDM one.}
\end{minipage}
\end{figure}

We find a local minimum of $\chi^2$ for  $\Omega_{dm}=0.345865$ and $\Omega_{de}=0.606504$ ($\chi^2=8.69263$) and present  the observational $H(z)$ data in Fig. 1 with error bars and the theoretical line corresponding to the best fit parameter.
The Fig. 2 shows that the Universe begins to accelerate on $z \sim 0.66$. A similar result can be obtained from  $\Lambda$CDM flat cosmology when the density  parameters are $\Omega_m=0.3$ and $\Omega_{\Lambda}=0.7$ \cite{Copeland:2006wr}.
In Fig. 3 we plot confidence regions in the $\om_{dm}$ - $\om_{de}$ plane. The true values of those parameters are inside the inner ellipse or between both ellipses with $68.3$ or $95.4$ percent of probability respectively. In Fig. 4 we plot the equations of state for dark matter $w_{dm}=p_{dm}/\rho_{dm}$, dark energy $w_{de}=p_{de}/\rho_{de}$ and overall fluid $w=p/\rho$. According to the previous stability analysis predictions we see that the asymptotic behavior of the overall state parameter varies from $ w\sim 0$ (near cold matter behavior) in the far past to $ w=-1+A^2/3\sim -0.63$ in the far future (dark energy behavior). In Fig. 5 we show the energy densities  corresponding to the baryonic and dark components in terms of the  redshift $z$. 

We calculate the age of the universe (\ref{t0}) with the best fit parameters and find that $t_0=19.885 Gyr$. In Fig. 6 we plot the time elapsed (in units of $H_0^{-1}$), since the initial singularity to present days, for our model and the flat  $\Lambda$CDM model, as a function of the matter density. We also show the border $t_0 = 11 Gyr$ coming from the bound of the oldest stellar ages. The age of the universe in this coupled scenario tends to be much higher when compared with the ACDM case \cite{Franca:2003zg}

\section{Linear perturbations}

Cosmological models with two interacting fluids have been investigated with the purpose of describe the evolution of dark components. There the energy momentum tensor of the interacting components is not separately conserved. Usually these cosmological model are presented with interacting matter species have a non constant equation of state parameter \cite{koi} or with DE having a constant equation of state parameter coupled to DM \cite{ladw},\cite{diego}. However, in the model we are investigating the interaction between the two fluids is setting by Eq. (\ref{c2}). For this choice the field equation can be integrated, generalizing the case of quintessence driven by the exponential potential. Then, it will be interesting to investigate the evolution of the density perturbation.

In the synchronous gauge the line element is given by: 
\be
\n{ds}
ds^2 = a^2(\tau)[-d\tau^2 + (\delta_{ij} + h_{ij})dx^idx^j], 
\ee
where the commoving coordinate are related to the proper time $t$ and position ${\bi r}$ by $d\tau=dt/a$, $d{\bi x}=d{\bi r}/a$ and $h_{ij}$ is the metric perturbation.
The scalar mode of $h_{ij}$ is described by the two fields $h({\bi k},\tau)$ and $\eta(\bi k,\tau)$ in the Fourier space, 
\be
\n{def}
h_{ij}(\bi x,\tau)=\int d^3 k e^{i\bi k{\bi\cdot}\bi x}\left[\hat{\bi k_i }\hat{\bi k_j} h + (\hat{\bi k_i }\hat{\bi k_j} -\frac{1}{3}\delta_{ij})\eta \right]. 
\ee
with ${\bi k} = k \hat{\bi k}$. The Einstein equations to linear order in
k-space, expressed in terms of $h$ and $\eta$, are given by the following
four equations \cite{Ma}:
\ben                                                                  
\n{a1}
&&k^2\eta - \frac{1}{2}\frac{a'}{a}h'= 4\pi G a^2 \delta T^0_0, \\
\n{a2}
&&k^2 \eta' = 4\pi G a^2 (\rho + p)\theta,   \\
\n{a3}
&&h'' + 2\frac{a'}{a}h' -2k^2\eta = -8\pi G a^2 \delta T^i_i,  \\  
\nonumber
&&h''+6\eta''+2\frac{a'}{a}(h'+6\eta') -2k^2\eta =\\
\n{a4}
&&-24\pi G a^2 (\rho + p)\sigma.
\een
Here, the quantities $\theta $ and $\sigma$ are defined as $(\rho + p)\theta = i k^j \delta T^0_j$, $(\rho + p)\sigma = -(\bi k_i\bi k_j - \delta_{ij}/3)\Sigma^i_j$ and $\Sigma^i_j = T^i_j - \delta^i_j T^k_k/3$ denotes the traceless component of the tensor $T^i_j$. In addition,  $\theta $ is the divergence of the fluid velocity $\theta = i k^jv_j$ and $'$ means $d/d\tau$.

Let us consider a fluid moving with a small coordinate velocity $v^i = dx^i/d\tau$, then, $v^i$ can be treated as a perturbation of the same order as energy density, pressure and metric perturbations. Hence, to linear order in the perturbations, the energy-momentum tensor, with vanishing anisotropic shear perturbation $\Sigma^i_j$, is given by
\ben
\n{tpert1}
T^0_0 = - (\rho + \delta\rho), \\
\n{tpert2}
T^0_i =  (\rho + p)v_i = - T^i_0, \\
\n{tpert3}
T^i_j =  (p + \delta p)\delta^i_j.
\een

For a fluid with equation of state $p=w\rho$, the perturbed part of energy-momentum conservation equations $T^{\mu\nu}_{;\mu} = 0$ in the k-space leads to the equations
\ben
\n{pertd}
\delta' = -(1+w)(\theta + \frac{h'}{2}) - 3\mathcal H (\frac{\delta p}{\delta \rho} - w)\delta,   \\
\n{tet}
\theta' = - \mathcal H (1 -3w)\theta - \frac{w'}{1+w}\theta
+ \frac{\delta p/\delta \rho}{1 + w}k^2\delta, 
\een
\no where $\delta = \delta\rho/\rho $ and $\mathcal H = a'/a=a H=\dot a$. Besides, using equations (\ref{a1}), (\ref{a3}), (\ref{tpert1}) and (\ref{tpert3}) we arrive at 
\be
\n{h}
h'' + \mathcal H h' + 3\mathcal H^2\left(1 +3\frac{\delta p}{\delta\rho}\right)\delta = 0.
\ee

We have showed that our interacting two-fluid model can be associated with an overall perfect fluid description based in an effective equation of state $w=(w_1 \rho_1+ w_2 \rho_2)/(\rho_1+\rho_2)$. Hence, we investigate the asymptotic regimes at early and late times assuming nearly constant equations of state $w\approx w^e=0$ and $w\approx w^l=-1+A^2/3$ respectively. 

At early time, when the overall fluid has $w\approx 0 $, the effective fluid perturbations evolve similar to those of ordinary dust with $\dot\theta=\theta=0$, and from Eqs. (\ref{pertd}-\ref{h}) we obtain
\be
\n{early1}
\ddot \delta + 2 H \dot\delta - \frac{3}{2}H^2\delta = 0
\ee 
and $\delta=c_1 t^{-1}+c_2 t^{2/3}$, where $c_1$ and $c_2$ are arbitrary integration constants. In this dust dominated era the perturbation grows as $\delta\approx a$ showing an initial unstable phase and compatible with the observation that the primordial universe would have tiny perturbations which seed the formation of structures in the later universe.

At late times, we are interested to find the evolution of the linear scalar perturbations for any mode $k$. To this end we write the second order differential equation for the density perturbation $\delta$ and the first order differential equation for the divergence of the fluid velocity $\theta$, evaluating them on the asymptotically stable equation of state $w\approx w^l$. In this case, from Eqs. (\ref{pertd}-\ref{h}) we get:
\ben
\nonumber
\delta''+\mathcal{H}\delta'+\left[w^l k^2-\frac{3}{2}(1+w^l)(1+ 3w^l)\mathcal{H}^2\right]\delta\\
\n{delt1} 
+ 3w^l(1+w^l)\mathcal{H}\theta=0,   \\
\n{delt2} 
\theta'=-\mathcal{H}(1-3w^l)\theta + \frac{w^l}{1+w^l}k^2\delta. 
\een 

Taking into account that in the late time regime the scale factor behaves as $a\propto t^{2/3(1+w^l)}$ we can calculate the conformal time $\tau$, $a$ and $\mathcal{H}=a'/a$
\ben
\n{eta}
&&\tau\propto t^{(1+3w^l)/3(1+w^l)}\\
\n{a}
&&a\propto \tau^{2/(1+3w^l)}\\
\n{.a}
&&\mathcal{H}=\frac{2}{(1+3w^l)\,\tau}.
\een
From Eqs. (\ref{delt1}) and (\ref{delt2}) the perturbation evolution becomes mode dependent with the $k^2/\mathcal{H}^2$ term, and for low energy modes their solutions can be obtained assuming a power law dependence of the perturbations with the scale factor, $\delta\propto a^n$ and $\theta\propto a^s$. In this case the approximate solutions for $w^l=-1+A^2/3=-0.63$ are given by
\ben
\n{st}
&&\theta\approx\frac{\theta_0}{ a^{2.89}}\\
\n{sd}
&&\delta\approx \frac{\delta_1}{a^{0.55}}+\frac{\delta_2}{a^{0.89}}+\frac{\theta_1}{a^{3.33}},
\een
where $\theta_0$, $\delta_1$ and $\delta_2$ are integration constants while $\theta_1$ is a function of $\theta_0$ and $w^l$. This shows that the coupling to $\theta$ in Eq. (\ref{delt1}) can be neglected for all scales we are interested. Finally, expressing the Eq. (\ref{delt1}) in term of conformal time we get
\be
\n{d2}
\delta''+\frac{2}{1+3w^l}\,\,\frac{\delta'}{\tau}+\left[wk^2-6\,\frac{1+w^l}{1+3w^l}\,\,\frac{1}{\tau^2}\right]\delta=0.
\ee
   The general solution of the latter equation in terms of the Bessel functions is
\be
\n{d3}
\delta=\tau^b\left[c_1J_\nu(k\sqrt{w^l}\tau)+c_2J_{-\nu}(k\sqrt{w^l}\tau)\right],
\ee
with
\be
\n{pnu}
b=\frac{-1+3w^l}{2(1+3w^l)}, \qquad  \nu=\pm\frac{5+9w^l}{2(1+3w^l)}.
\ee
At late times, it can be approximated by the two first terms of the Eq. (\ref{sd}) showing that the energy density perturbation decreases for large cosmological times for modes satisfying the condition $k^2/\mathcal{H}^2\ll 1$. For high energy modes, $k^2/\mathcal{H}^2\gg 1$, the perturbation 
\be 
\n{ma}
\delta\approx \frac{1}{a^{0.5}},
\ee
decreases but slowly that the low energy modes.
This results can be understood writing the Eq. (\ref{delt1}) as the
equation of motion of a dissipative mechanical system by using the analogy with the classical potential problem
\be
\n{em}
\frac{d}{d\tau}\left[\frac{\delta'^2}{2}+\mathcal{V}(\delta)\right]=-D(\delta,\delta'),
\ee
where 
\ben
\n{V}
&&\mathcal{V}(\delta)=w^lk^2\left(1-\frac{\mathcal{H}^2}{\mathcal{H}^2_0}\right)\frac{\delta^2}{2},\\
\n{D}
&&D(\delta,\delta')=\frac{3}{2}(1+w^l)(1+ 3w^l)\mathcal{H}\mathcal{H}'\delta^2+\mathcal{H}\delta'^2,\\
\n{h0}
&&\mathcal{H}^2_0=\frac{2w^l k^2}{3(1+w^l)(1+ 3w^l)}.
\een
The potential $\mathcal{V}$ has an extreme at $\delta=0$, it is maximum for $\mathcal{H}<\mathcal{H}_0$ or a minimum  for $\mathcal{H}>\mathcal{H}_0$. On the other hand, assuming that the perturbation depends on the scalar factor in the form $\delta\propto a^n$, we find that $D\approx 0.08\mathcal{H}^3\delta^2>0$. Hence, for any mode $k$ the perturbation begins to grow at early times for $\mathcal{H}<\mathcal{H}_0$, while at late times for $\mathcal{H}>\mathcal{H}_0$, the function inside the square bracket in Eq. (\ref{em}) is a Liapunov function and the perturbation decreases asymptotically reaching $\delta=0$ in the limit $t\to\infty$.

\section{CONCLUSIONS}

We have shown an interacting two-fluid cosmological model that allows us
to reproduce the accelerated behavior of our universe and its probable age .
The model gives rise to an exotic scalar field dubbed exotic quintessence
which reduces to quintessence  when one fluid is associated with stiff
matter and the other with vacuum energy. Setting the interaction between
the two fluids by Eq. (\ref{c2}), the field equation is integrated,
generalizing the case of quintessence driven by the exponential potential
and, the equation governing the scale factor (\ref{F}) looks like a
modified Friedmann equation.

We have obtained the evolution equation for the overall equation of state 
of the model and showed the asymptotic behavior of the scale factor i.e.,
the universe begins from an unstable phase dominated by the first
fluid, $a\approx t^{2/3(1+w_1)}$, and ends in a stable expanding phase
dominated by the exponential potential, $a\approx t^{2/A^2}$.
The latter becomes accelerated when the exponential potential slope
satisfy $A^2<2$. Setting $w_1=0$, the scale factor interpolates
between pressureless matter and dark energy phases.

	Using the Hubble function $H(z)$ data from Table 1 we minimize the $\chi^2$ function (\ref{ch1}), obtaining the best fit densities parameters  $\om_{dm}=0.346^{+0.054}_{-0.046}$ and $\om_{de}=0.606^{+0.014}_{-0.026}$  with a reduced $\chi^2=1.24$. 
 These results are consistent with those found in the literature, see  for instance Ref. \cite{Amendola:2006dg} for null coupling ($\om_{dm}=0.37^{+0.06}_{-0.08}$) or with the result obtained in Ref. \cite{Feldman:2003nu}, ($\om_{dm}=0.30^{+0.17}_{-0.07}$) through mean relative peculiar velocity measurements for pairs of galaxies. 
    With our best densities parameters and the priors for $H_0=72 km s^{-1} Mpc^{-1}$ and  $\om_b=0.05$, we obtain the theoretical $H(z)$ function, plotted together with the SVJ(2005) experimental data in Fig. 1. In Fig. 2 we plot the acceleration of the model as a function of the redshift and find the transition from the non accelerated phase to the  accelerated one around $z_{acc} = 0.66$ \cite{Copeland:2006wr}. Our value agrees with the result obtained by a nearly model independent characterization of dark energy properties as a function of redshift, ($z_{acc}=0.42^{+0.08}_{-0.06}$,\cite{Daly:2006ax}). The problem of why an accelerated expansion should occur now in the long history of the universe seems to be naturally dressed in our model.  
	Considering the age of the universe, we take into account that the age of the oldest stellar objects have been constrained for instance, by using a distance-independent method \cite{Jimenez:1996}, ($t_0=13.5 \pm 2Gyr$ for Globular clusters in the Milky Way) and  
the white dwarfs cooling sequence method \cite{Hansen:2002ij}  ($t=12.7\pm 0.7 Gyr$ for the globular cluster M4). Then, the age of universe needs to satisfy the lower bound $t_0 > 12 - 13 Gyr$.  This condition is fulfilled by our model with $t_0=19.885 Gyr$, as it can be seen in Fig. 5. 

The energy density perturbation of the model grows in the first stage of the universe  
showing that initial instabilities in the primordial universe could leads to the formation of structure in the later universe. At late times we have found a  Liapunov function which indicates that the perturbation decreases asymptotically reaching $\delta=0$ in the limit $t\to\infty$.

\section*{Acknowledgments}

This work was partially supported by the University of Buenos Aires and
Consejo Nacional de Investigaciones Cient\'\i ficas y T\'ecnicas under
Projects X224 and 5169.


\end{document}